\documentclass[%
 reprint,
 amsmath,amssymb,
 aps,
prx,
floatfix,
]{revtex4-1}

\usepackage{ifpdf}
\usepackage{amsmath} 
\usepackage{amssymb}
\usepackage{amsfonts}
\usepackage{braket}
\usepackage{physics}
\usepackage{color}
\usepackage{booktabs}
\usepackage[colorlinks=true,linkcolor=blue,citecolor=blue,breaklinks]{hyperref}
\usepackage{graphicx}
\usepackage{dcolumn}
\usepackage{bm}
\usepackage[normalem]{ulem}
\usepackage{mathtools}
\usepackage{verbatim}
\usepackage{xcolor}

\newcommand{\hc}{\text{H.c.}}

\usepackage{tikz}
\usetikzlibrary{math}

\definecolor{Zcolour}{rgb}{0.992, 0.588, 0.22}
\definecolor{purple}{rgb}{0.5,0,0.5}
\definecolor{brown}{rgb}{0.6,0.2,0}
\definecolor{dkgreen}{rgb}{0,0.5,0}

\begin{document}


\title{New exact eigenstates in the Lesanovsky model, proximity to integrability and the PXP model, and approximate scar states}
\author{Daniel K.~Mark}
\affiliation{Department of Physics, California Institute of Technology, Pasadena, CA 91125, USA}
\author{Cheng-Ju Lin}
\affiliation{Perimeter Institute for Theoretical Physics, Waterloo, Ontario N2L 2Y5, Canada}
\author{Olexei I.~Motrunich}
\affiliation{Department of Physics, California Institute of Technology, Pasadena, California 91125, USA}

\date{\today}

\begin{abstract}

We study a model of Rydberg atoms in a nearest-neighbor Rydberg blockaded regime, introduced by Lesanovsky in Phys.~Rev.~Lett.~{\bf 108}, 105301 (2012). This many-body model (which has one parameter $z$) has an exactly known gapped liquid ground state, and two exactly known low-lying excitations. We discover two new exact low-lying eigenstates.  We also discuss behavior of the model at small parameter $z$ and its proximity to an integrable model. Lastly, we discuss connections between the Lesanovsky model at intermediate $z$ and so-called PXP model. The PXP model describes a recent experiment that observed unusual revivals from a charge density wave initial state, which are attributed to a set of many-body ``scar states" which do not obey the eigenstate thermalization hypothesis. We discuss the possibility of approximate scar states in the Lesanovsky model and present two approximations for them.

\end{abstract}

\maketitle

\section{Introduction} \label{section:Intro}

Cold Rydberg atoms have received significant recent attention.  These are experimentally realizable systems of atoms trapped in optical lattices~\cite{jaksch_fast_2000,weimer_rydberg_2010}, which can be manipulated for a variety of purposes, including quantum simulation~\cite{schaus_observation_2012,pritchard_cooperative_2010,Bernien2017,celi_emerging_2019} and quantum computing~\cite{wilk_entanglement_2010,isenhower_demonstration_2010,saffman_quantum_2016}.

Rydberg systems consist of atoms trapped at optical lattice sites $j$, with valence electrons in their atomic ground state denoted $\ket{0}_j$.  A highly excited Rydberg state $\ket{1}_j$ is made accessible by Rabi oscillations from a driving laser.  Strong repulsive van der Waals interactions between Rydberg states lead to the possibility of engineering strongly interacting quantum systems.

In particular, the distance between Rydberg atoms can be tuned such that the nearest neighbor interactions between Rydberg states are effectively infinite, while longer range interactions are negligible.  This nearest-neighbor Rydberg blockade regime results in a constrained Hilbert space:  
The states of a one-dimensional (1d) chain of $L$ Rydberg atoms can be written in a product state basis of binary strings of $L$ `0's and `1's, subject to the condition that no two `1's can be next to each other.  
The dimension of the Hilbert space grows as $\sim \phi^L$, where $\phi = (1+\sqrt{5})/2 \approx 1.618$ is the golden ratio~\cite{LesanovskyPhysRevA.86.041601,Turner2018}.

Some time ago, Lesanovsky~\cite{Lesanovsky2012} studied a family of Hamiltonians in this constrained Hilbert space.  Hamiltonians in this family possess an exactly known gapped liquid ground state.  Lesanovsky also found two low-lying exact states, one of which is the first excited state, at finite energy above the ground state.

A recent experiment~\cite{Bernien2017} brings renewed interest in the Lesanovsky model.  The experiment, conducted in the same nearest-neighbor blockade regime, observed unusual quench dynamics of a $\ket{\mathbb{Z}_2} \equiv \ket{\dots 101010 \dots}$ charge density wave (CDW) initial state, dubbed the $\mathbb{Z}_2$ state.  Subsequent theoretical analysis~\cite{Turner2018} modelled the system with the so-called ``PXP model," and attributed the unusual dynamics to the presence of ``many-body scar states," a set of exceptional eigenstates in the many-body spectrum with unusually high overlap with the CDW $\mathbb{Z}_2$ state.  These scar states are of interest because they violate the strong eigenstate thermalization hypothesis (ETH).

The ETH has emerged as a paradigm for thermalization in closed quantum many-body systems~\cite{Deutsch1991, Srednicki1994}.  A \emph{strong} ETH appears to hold in many systems, where \emph{all} eigenstates at finite energy density obey the ETH. Many-body localized systems~\cite{Basko2006, serbyn_local_2013,nandkishore_many-body_2015} strongly violate the ETH. However, the PXP model belongs to a group of systems~\cite{shiraishi_systematic_2017, Moudgalya2018, moudgalya_entanglement_2018,moudgalya_quantum_2019, bull_systematic_2019, schecter_weak_2019,iadecola_quantum_2019,khemani_local_2019,chattopadhyay_quantum_2019,hudomal_quantum_2019,pancotti_quantum_2019} where the \emph{weak} ETH holds; that is, the ETH holds for \emph{almost} every eigenstate.

While numerous studies have shed insight into the PXP model~\cite{Turner2018,turner_quantum_2018,Iadecola2019,khemani_signatures_2018,ho_periodic_2019,michailidis_slow_2019, Choi2019, surace_lattice_2019,Lin2019,lin_slow_2019}, analytical understanding remains wanting. In particular, only two states are exactly known in periodic boundary conditions (PBC)~\cite{Lin2019}.
The Lesanovsky model is related to the PXP model and offers an attractive alternative for analysis.

\begin{figure*}[tb]
\centering
\includegraphics[width=\textwidth]{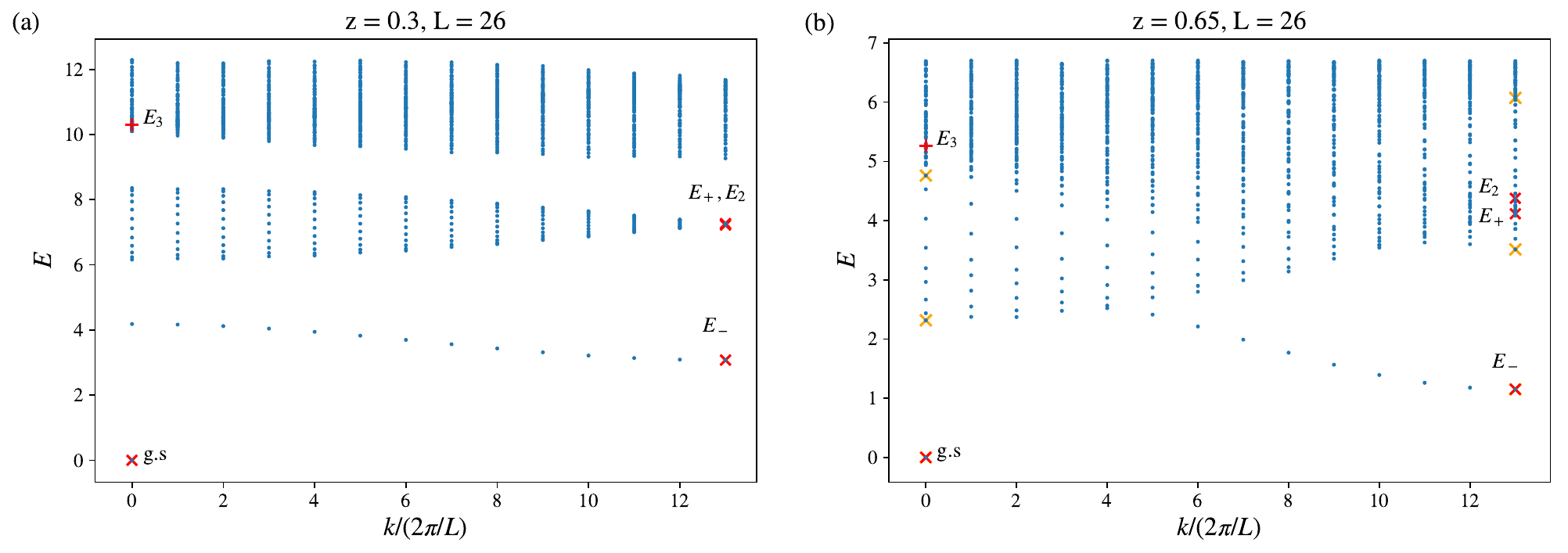}
\caption{Dispersion of eigenstates in the Lesanovsky model for (a) $z=0.3$ and (b) $z=0.65$. For (a), the first 4 ``bands'' of states are plotted, while for (b) all bands overlap and the same number of states are plotted. Exact states $\ket{z}, \ket{E_\pm}, \ket{E_2}$ are marked with red crosses `$\times$'. The point $k=0, E = 3 z^{-1} + z$ is marked with a `+' to indicate $\ket{E_3}$. This state is only present as an exact eigenstate for odd $L$ and is labelled here for reference only. Note that in (a) $E_2 - E_+ \approx 0.043$, and the two crosses overlap. The orange crosses in (b) mark the $\mathbb{Z}_2$ outlier (``scar") states discussed in Section \ref{section: PXP model}.}
\label{fig: dispersion}
\end{figure*}

In this paper, we revisit the Lesanovsky model and obtain several new results, presented as follows. In Sec.~\ref{section:Model} we review the Lesanovsky model, which has one parameter $z$, and its analytical ground state $\ket{z}$ and the first excited state $\ket{E_-}$. In Sec.~\ref{section:New states} we present two new exact eigenstates of the Lesanovsky model, $\ket{E_2}$ and $\ket{E_3}$, which we will define later in Eqs.~(\ref{eq: E2}) and (\ref{eq: E3}). These states and previously known exact states in the Lesanovsky model are summarized in Table~\ref{tab: states} and labelled in Fig.~\ref{fig: dispersion}.
In Sec.~\ref{section:Schrieffer-Wolff} we then discuss the Lesanovsky model in the low-$z$ limit. In particular, through a Schrieffer-Wolff transformation working perturbatively in small $z$, we show that the Lesanovsky model approaches an integrable model as $z \rightarrow 0$. This gives us a better physical picture of the excitations at low $z$. Lastly, in Sec.~\ref{section: PXP model} we discuss the relationship between the PXP model and the Lesanovsky model, in particular for $z\approx 0.65$, which is the parameter we found where the two models are ``closest" to each other. 
We discuss approximate scar-like states in the Lesanovsky model in this regime and present two approximations of them.
These approximations connect the ``scar states" to multi-quasiparticle states built out of the exactly known $\ket{E_-}$ excitation and to states in the aforementioned integrable model.

\section{Model} \label{section:Model}

We consider the following 1d model previously studied by Lesanovsky~\cite{Lesanovsky2012}:
\begin{equation} \label{eq:Hamiltonian}
    H = \sum_{j=1}^L P_{j-1} (X_j + z^{-1} n_j + z P_j) P_{j+1} ~,
\end{equation}
where $n_j = \ketbra{1}{1}_j$ is the Rydberg excitation occupation number, $P_j = \ketbra{0}{0}_j$ is the ground state occupation number (equivalently, projector onto the ground state), $X_j = \ketbra{1}{0}_j + \ketbra{0}{1}_j$ is the tunnelling operator between the two atomic states, and $z$ is a parameter that can range from $0$ to $\infty$. In this paper, we take periodic boundary conditions (PBC), i.e., $P_0 = P_{L}$ and $P_{L+1} = P_{1}$.

The PXP model $H_{\text{PXP}}$ shares the PXP term with the Lesanovsky model, but does not have the chemical potential PnP and PPP terms.

The Lesanovsky and PXP models are both invariant under translation symmetry $j \to j+1$ and inversion symmetry $I_{\text{site}}: j \to -j$.  They are also invariant under time reversal symmetry given by complex conjugation in the occupation number basis.

Unlike the PXP model, the Lesanovsky model is not particle-hole symmetric. Here particle-hole symmetry refers to the fact that $\mathcal{C} H_\text{PXP} \mathcal{C} = - H_\text{PXP}$, where $\mathcal{C} = \prod_{j=1}^L Z_j$~\cite{Turner2018}, which implies that the PXP spectrum is symmetric about 0.

The ground state of the Lesanovsky model $\ket{z}$ is known exactly and has energy $0$:
\begin{align} \label{eq:Ground state}
    \ket{z} &= \sum_{\{\sigma_i \},~\text{Rydb.}} (-z)^{\sum_j n_j} \ket{\{\sigma_i \}} \\
    &= \exp(-z \sum_{j = 1}^L P_{j-1} \sigma_j^+ P_{j+1}) \ket{00...0} ~,
\end{align}
where the sum in the first line is taken over all strings $\{\sigma_i \}$ satisfying the Rydberg blockade. In the second line, $\sigma_j^+ = \ketbra{1}{0}_j$, and the different operators in the sum in the exponent all commute. Here and subsequently we omit normalization factors of wavefunctions. Every allowed product state in the Hilbert space occurs in $\ket{z}$, with each `1' weighted by a factor of $(-z)$.

Additionally, for $L$ even, Lesanovsky found two other exact low-lying states $\ket{E_\pm}$, with $\ket{E_-}$ being the first excited state~\cite{Lesanovsky2012}:
\begin{equation} \label{eq:Lesanovsky states}
    \ket{E_\pm} = \sum_{j=1}^L (-1)^j [\alpha_\pm n_j + \beta n_{j-1} n_{j+1}] \ket{z} ~,
\end{equation}
where $E_\pm$ and $\alpha_\pm, \beta$ satisfy the eigenvalue equation:
\begin{equation} \label{eq: Lesanovsky states equation}
    \begin{pmatrix}
    z^{-1} - z & 2z \\
    -z & 2z^{-1} + 2z
    \end{pmatrix}
    \begin{pmatrix}
    \alpha_\pm \\
    \beta
    \end{pmatrix} = E_\pm        
    \begin{pmatrix}
    \alpha_\pm \\
    \beta
    \end{pmatrix} ~. 
\end{equation}

\begin{table}[tb] 
\centering
\begin{tabular}{c | c c  c c}
\textbf{Eigenst.} & $E$ & $k$ & $I_\text{site}$ & $L$\\
\hline
$\ket{z}$ & $0$   & $0$ & 1 & all \\
$\ket{E_-}$ & $(3+z^2 - \sqrt{1+6z^2+z^4})/(2z)$ & $~\pi~$ & 1 & ~even~ \\
$\ket{E_+}$ & $(3+z^2 + \sqrt{1+6z^2+z^4})/(2z)$ & $\pi$ & 1 & even\\
$\ket{E_2}$ & $2z^{-1} + 2z$ & $\pi$ & -1  & even\\
$\ket{E_3}$ & $3z^{-1} + z$ & $0$ & -1 & odd\\
\end{tabular}
\caption{Exact eigenstates known in the Lesanovsky model with periodic boundary conditions (PBC) and their quantum numbers. The new states in this work $\ket{E_2}$ and $\ket{E_3}$ are introduced in Sec.~\ref{section:New states}}
\label{tab: states}
\end{table}

Lastly, we point out an interesting relation between the Lesanovsky model and the Fendley-Sengupta-Sachdev (FSS) model studied in Ref.~\cite{Fendley2004}. Because the PPP term in the constrained Hilbert space can be rewritten as $\sum_j P_{j-1} P_j P_{j+1} = L + \sum_j n_{j-1} n_{j+1} - 3n_j$, the negative of the
Lesanovsky model defines a curve in the FSS family of models~\cite{Chepiga2019, chepiga_floating_2019}. Hence, the ``ceiling state" (i.e., the highest energy state) of the Lesanovsky model corresponds to the ground state of the FSS model and undergoes an Ising second order transition from a $\mathbb{Z}_2$ ordered phase at low $z$ to a disordered phase at high $z$. In the thermodynamic limit, when the ceiling state transitions from disordered to $\mathbb{Z}_2$ ordered, the gap between the ceiling states at $k=0$ and $k=\pi$ closes. While the gap is never 0 for finite systems, as we approach the transition from high $z$, the energy gap decreases linearly with $z$, with deviations from linearity only near the transition itself. By interpolating this linear trend, we numerically determined this transition to occur at $z \approx 0.49$. When $z<0$, the state $\ket{z}$ in Eq.~(\ref{eq:Ground state}) is in fact the ceiling state of the Lesanovsky model, and the Lesanovsky model corresponds to the `disorder line' of the FSS model, separating commensurate and incommensurate regions of the disordered phase~\cite{Chepiga2019,Fendley2004}.

\section{New Exact States in the Lesanovsky Model} \label{section:New states}
In addition to the exact states $\ket{z}, \ket{E_{\pm}}$ introduced in Ref.~\cite{Lesanovsky2012}, we discovered additional exact low-lying states $\ket{E_2}$ and $\ket{E_3}$. By adopting the approach of Refs.~\cite{Moudgalya2018, Lin2019} and examining the Schmidt numbers of eigenstates obtained from exact diagonalization (ED) of finite systems, we found two new states with finite Schmidt number of 12 and 16, indicating these states might be ``simple" and hinting at their exact expressions. $\ket{E_2}$ has Schmidt number 12 for all even $L\geq12$, and $\ket{E_3}$ has Schmidt number 16 for all odd $L\geq17$. Apart from the known states $\ket{z}$ and $\ket{E_\pm}$ (with Schmidt numbers 4 for $L\geq6$ and 8 for even $L\geq10$ respectively), these were the only other states observed to have finite Schmidt number.

We present the proofs of the $\ket{E_2}$ and $\ket{E_3}$ states in Sec.~\ref{subsection: Proofs}, by studying a transformed Hamiltonian Eq.~(\ref{eq: rotated Ham}).

\subsection{Exact eigenstate $\ket{E_2}$ in even $L$}
By analyzing the eigenstate with Schmidt number 12, we found it has energy $E = 2z^{-1} + 2z$ and has (unnormalized) expression:
\begin{equation} \label{eq: E2}
    \ket{E_2} = \sum_{j=1}^L (-1)^j n_j n_{j+3} \ket{z} ~.
\end{equation}
$\ket{E_2}$ carries momentum $\pi$ and has inversion quantum number $I_{\text{site}}=-1$. 
It can be thought of as a `1001' bound state excitation at wavevector $\pi$  on top of the ground state $\ket{z}$.

\subsection{Exact eigenstate $\ket{E_3}$ in odd $L$}
$\ket{E_3}$ is present on odd-$L$ chains, with energy $E = 3z^{-1} + z$, and has (unnormalized) expression:
\begin{equation} \label{eq: E3}
    \ket{E_3} = \sum_{j=1}^L \sum_{k=1}^L (-1)^k n_{j-1} n_{j+1} n_{j+k} \ket{z} ~.
\end{equation}
Some care must be taken with the factor $(-1)^k$: $k$ must be taken strictly between 1 and $L$, because $L$ is odd. For example, for a fixed $j$, though $k = -1$ and $k = L-1$ would give the same site $j+k$, these $k$ values would give opposite signs: $(-1)^{-1} = - (-1)^{L-1}$; it is the latter $k = L-1$ that we use.

This state can be thought of as a scattering state of a `1' excitation and a bound excitation `101' on top of the ground state. Interestingly, the `1' and `101' particles appear as if they have mutual statistics of $-1$: exchanging the position of the `1' and `101' particles gives a relative phase of $-1$. This gives the state an inversion number of $I_\text{site} = -1$. This state can be equivalently thought of as a superposition of Lesanovsky $E_-$ and $E_+$ particles that is antisymmetric under exchange.

\subsection{Rotated Hamiltonian and proofs of eigenstates} \label{subsection: Proofs}

These states can be proven by performing the following similarity transformation (conjugation) of the Lesanovsky model:
\begin{align}
 H_R &= S^{-1} H S ~, \quad S = \exp(-z \sum_{j=1}^L P_{j-1} \sigma_j^+ P_{j+1}) ~.
\end{align}
Straightforward algebra gives
\begin{align}
 H_R = \sum_{j=1}^L \bigg[& (z^{-1} + z) n_j + z P _{j-1} (\sigma_j^+ \sigma_{j+1}^- + \sigma_j^- \sigma_{j+1}^+) P_{j+2} \nonumber \\
 & + \sigma_j^- + z^2 P_{j-2} \sigma_{j-1}^+ \sigma_{j}^-\sigma_{j+1}^+ P_{j+2} \bigg] ~.   
 \label{eq: rotated Ham}
\end{align}

The rotated Hamiltonian $H_R$ is non-Hermitian because the conjugation operator $S$ is non-unitary. Nevertheless, $H_R$ and $H$ have the same eigenvalues, and the corresponding eigenstates are related by $S$. $H_R$ has two terms that conserve the number of `1's (Rydberg excitation number): a chemical potential $n_j$ term, and a hopping $P(\sigma^+ \sigma^- + \hc)P$ term. It also has a term $\sigma_j^-$ that lowers the excitation number by sending `1's to `0's, and a term $P \sigma^+ \sigma^- \sigma^+ P$ that raises the excitation number by sending `1's to `101's.

In this frame, the Lesanovsky ground state from Eq.~(\ref{eq:Ground state}) is simply $\ket{0...0}$. The Lesanovsky excited states from Eq.~(\ref{eq:Lesanovsky states}) are also simply linear combinations of `1' magnon states $\sum_j (-1)^j \ket{0...01_j0...0}$ and `101' ``bound magnon" states $\sum_j (-1)^j \ket{0...10_j1...0}$. 
This can be seen by noting that $n_j \ket{z} = -z P_{j-1} \sigma_j^+ P_{j+1} \ket{z}$, and that $P_{j-1} \sigma_j^+ P_{j+1}$ and $P_{k-1} \sigma_k^+ P_{k+1}$ commute for all $j,k$.

Similarly, the (unnormalized) states $\ket{E_2}$ and $\ket{E_3}$ in this frame are:
\begin{align}
    \label{eq: E2 rot}
    \ket{E_2}_R &= \sum_{j=1}^L (-1)^j \sigma^+_j \sigma^+_{j+3} \ket{0...0} \\
    &= \sum_{j=1}^L (-1)^j \ket{0...1_j001...0}, \nonumber\\
    \label{eq: E3 rot}
    \ket{E_3}_R &= \sum_{j=1}^L \sum_{k=1}^L (-1)^{k} \sigma_{j-1}^+ \sigma_{j+1}^+ \sigma_{j+k}^+ \ket{0...0} \\
    & = \sum_{j=1}^L \sum_{k=3}^{L-3} (-1)^k \ket{0...10_j1...1_{j+k}...0} ~. \nonumber
\end{align}

\textbf{Proof of $\ket{E_2}$:}
Due to the $k=\pi$ construction, $\ket{E_2}_R$ is killed by the hopping, lowering, and raising terms, and is trivially an eigenstate of the chemical potential term, with energy $E_2 = 2z^{-1} + 2z$.

\textbf{Proof of $\ket{E_3}$:}
Acting on $\ket{E_3}_R$, the raising and lowering operators will produce strings like $\ket{0...101...101...0}$ and $\ket{0...1...1..0}$ respectively, which cannot carry inversion number $I_\text{site} = -1$ because the strings are inversion symmetric. Therefore the sum of all contributions from such operators in $H_R$ is zero.

Under the hopping term, hopping of the `101' bound particle cancels, while hopping of the `1' particle gives energy $-2z$. The chemical potential term gives energy $3(z^{-1} + z)$, giving a total energy $E_3 = 3z^{-1} + z$.

\subsection{Additional approximate eigenstates}
The rotated Hamiltonian $H_R$ allows us to approximate several low-lying states in the spectrum. In particular, we study the $H_R$ matrix elements
between states corresponding to $\ket{E_\pm}$ and $\ket{E_2}$ at wavenumbers near $k=\pi$ in Appendix~\ref{section: Momentum ansatz}. These produce trial states that well approximate ED states near $k=\pi$. In particular, applying this method for odd length chains yields good approximates of the exact states $\ket{E_\pm}$ and $\ket{E_2}$ in odd system sizes. We also study $H_R$ mappings among states of low excitation number with zero total momentum in Appendix~\ref{section:Low energy integrable}. This inspires our `integrable ansatz' of scar states in the Lesanovsky model in Sec.~\ref{section:integrable ansatz}, dubbed `integrable' because of its connection to the integrable model discussed in Sec.~\ref{section:Integrable model}.

\section{Low $z$ effective Hamiltonian and physical pictures of excitations} \label{section:Schrieffer-Wolff}
For low $z$, the $z^{-1} \sum_j n_j$ term in $H$ dominates, and we get approximate sectors labelled by the total Rydberg excitation number as seen in Fig. \ref{fig:z 0.3 EE}. We can study the dynamics inside the sectors by performing a perturbative Schrieffer-Wolff type unitary transformation on the Hamiltonian described below. The primary difference between the resulting effective Hamiltonian and the rotated Hamiltonian $H_R$ discussed previously is that this effective Hamiltonian is Hermitian but truncated to some order in $z$, while the rotated Hamiltonian is non-Hermitian but exactly equivalent to the Lesanovsky model.

\subsection{Schrieffer-Wolff transformation}
We perform a Schrieffer-Wolff transformation to remove the $PXP$ term that mixes the sectors and obtain:
\begin{align}
    H_{\text{eff}}^{(0)} &= W^\dagger H W = \left(z^{-1} + z \right) \sum_j n_j     \label{eq:H0eff}\\
    &+ z \sum_j P_{j-1} (\sigma_j^+ \sigma_{j+1}^- + \sigma_j^- \sigma_{j+1}^+) P_{j+2} + O(z^2) ~, \nonumber
\end{align}
where 
\begin{equation}
    W = \exp\left[-z \sum_j P_{j-1} (\sigma_j^+ - \sigma_j^-) P_{j+1} \right] ~.
\end{equation}

\subsection{Proximity to an integrable model}
\label{section:Integrable model}
\begin{figure*}[tbh]
\centering
\includegraphics[width=\textwidth]{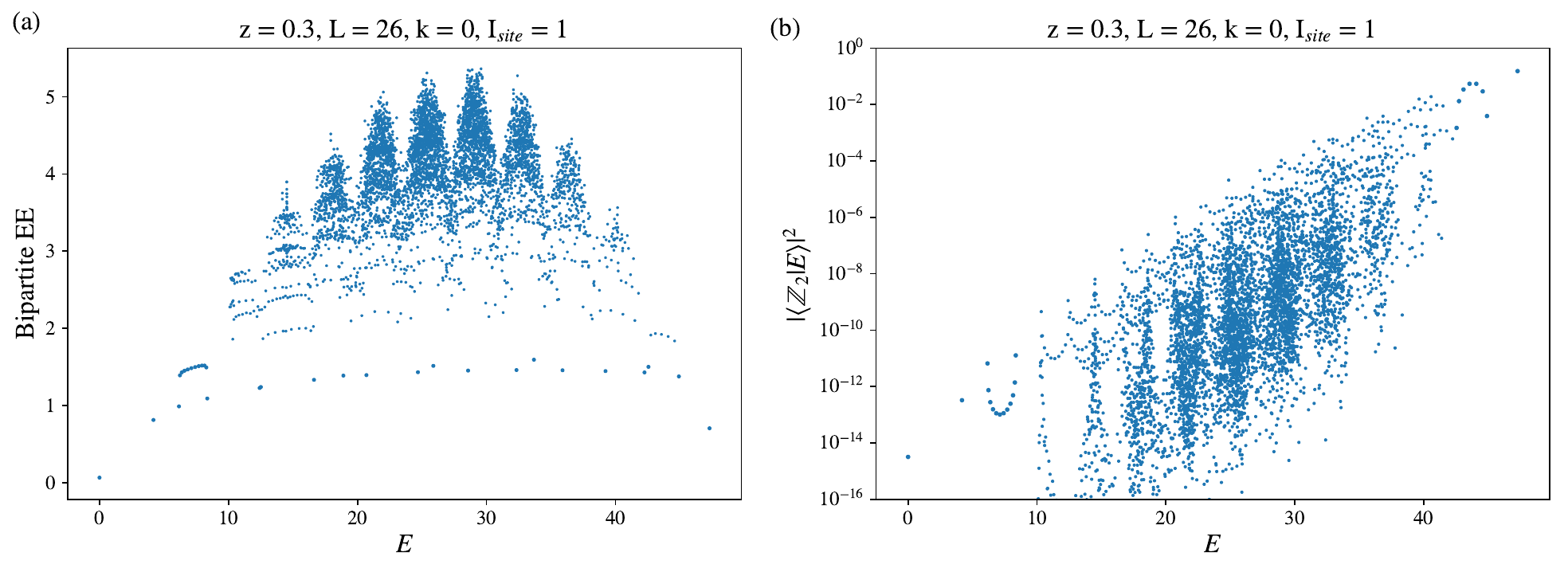}
\caption{(a) Bipartite entanglement entropy (EE) and (b) overlaps with the $\mathbb{Z}_2$ CDW at $z=0.3$, $L=26$, for all eigenstates with quantum numbers $k=0$ and $I_\text{site} = 1$. Note clear sectored structure. Different ``sectors" correspond to approximately conserved effective total Rydberg excitation number running from $0$ to $L/2$; the significant spread in the EE within each sector is due to proximity to integrability in the effective model. In (b), note that the ceiling state is in the $\mathbb{Z}_2$ ordered phase and has high ($\approx 0.15$) overlap squared with $\ket{\mathbb{Z}_2}$.
Also note the different character from Fig.~\ref{fig:z 0.65 Z2 overlap}, which indicates that the Lesanovsky model at $z=0.65$ is perturbatively far from the integrable model in Sec.~\ref{section:Integrable model}.}
\label{fig:z 0.3 EE}
\end{figure*}

$H_{\text{eff}}^{(0)}$ is in fact an integrable Hamiltonian described in Ref.~\cite{Alcaraz1999}, with Bethe ansatz solution. Eigenstates of $H_{\text{eff}}^{(0)}$ can be written as $n$ non-interacting magnons, each carrying some (possibly non-physical) momentum (``quasimomentum").

We can do further Schrieffer-Wolff transformations~\footnote{$H_{\text{eff}}^{(1)} = W_2^\dagger W_1^\dagger W^\dagger H W W_1 W_2$, where \\ $\begin{aligned}
    W_1 = &\exp\!\Big[-\frac{z^3}{3}\sum_j\Big( 4P_{j-2}\sigma^+_{j-1}\sigma^-_j \sigma^+_{j+1} P_{j+2} \\
    &- P_{j-1}\sigma^+_j P_{j+1}- P_{j-2}P_{j-1} \sigma^+_{j}P_{j+1} \\
    &- P_{j-1} \sigma^+_{j}P_{j+1}P_{j+2} - \hc\Big) \Big]\\
    \text{and}\\ 
    W_2 = &\exp \left[\frac{z^4}{2} \sum_j \left( P_{j-2}\sigma^+_{j-1}P_j\sigma^+_{j+1}P_{j+2} - \hc \right) \right]~. \end{aligned}$} to remove further sector mixing terms up to $O(z^4)$:
\begin{align}
    &H_{\text{eff}}^{(1)} = H_{\text{eff}}^{(0)} +\frac{1}{2}z^3 \Big(2P_{j-1}\sigma^+_j \sigma^-_{j+1}\sigma^+_{j+2}\sigma^-_{j+3}P_{j+4}\\
 &- P_{j-2}P_{j-1}\sigma^+_{j}\sigma^-_{j+1}P_{j+2}- P_{j-1}\sigma^+_{j}\sigma^-_{j+1}P_{j+2}P_{j+3} \nonumber\\
 & + \text{H.c.} \Big) + O(z^4) ~. \nonumber
\end{align}

The $z^3$ term breaks integrability of the effective Hamiltonian. At low $z$, the bipartite entanglement entropy (EE) has a sectored, nearly-integrable structure illustrated in Fig.~\ref{fig:z 0.3 EE}(a) for $z=0.3$. The sectors merge and thermalize as $z$ is increased, as shown in Fig.~\ref{fig:z 0.65 EE} for $z=0.65$.

\subsection{Describing exact eigenstates}
The eigenstates $\ket{E_\pm}, \ket{E_2}, \ket{E_3}$ in the effective Hamiltonian essentially agree with their descriptions in the rotated Hamiltonian in Sec.~\ref{subsection: Proofs}, such as
Eqs.~(\ref{eq: E2 rot}), (\ref{eq: E3 rot}) for the last two states.

$\ket{E_-}$ lies at the bottom of the 1-magnon band, cf.\ Fig.~\ref{fig: dispersion}(a). To order $z^3$, the 1-magnon band has dispersion $E(k) = z^{-1} + z + 2(z-z^3) \cos(k) + O(z^4)$, which at $k=\pi$ agrees with $E_-(z)$ expanded to this order.

$\ket{E_+}$ and $\ket{E_2}$ lie near the middle of the 2-magnon band. The degeneracy between $\ket{E_+}$ and $\ket{E_2}$ is broken by the $z^3$ term. 

Finally, $\ket{E_3}$ lies near (but not exactly at) the bottom of the 3-magnon band.

\begin{figure*}[bht]
\centering
\includegraphics[width=\textwidth]{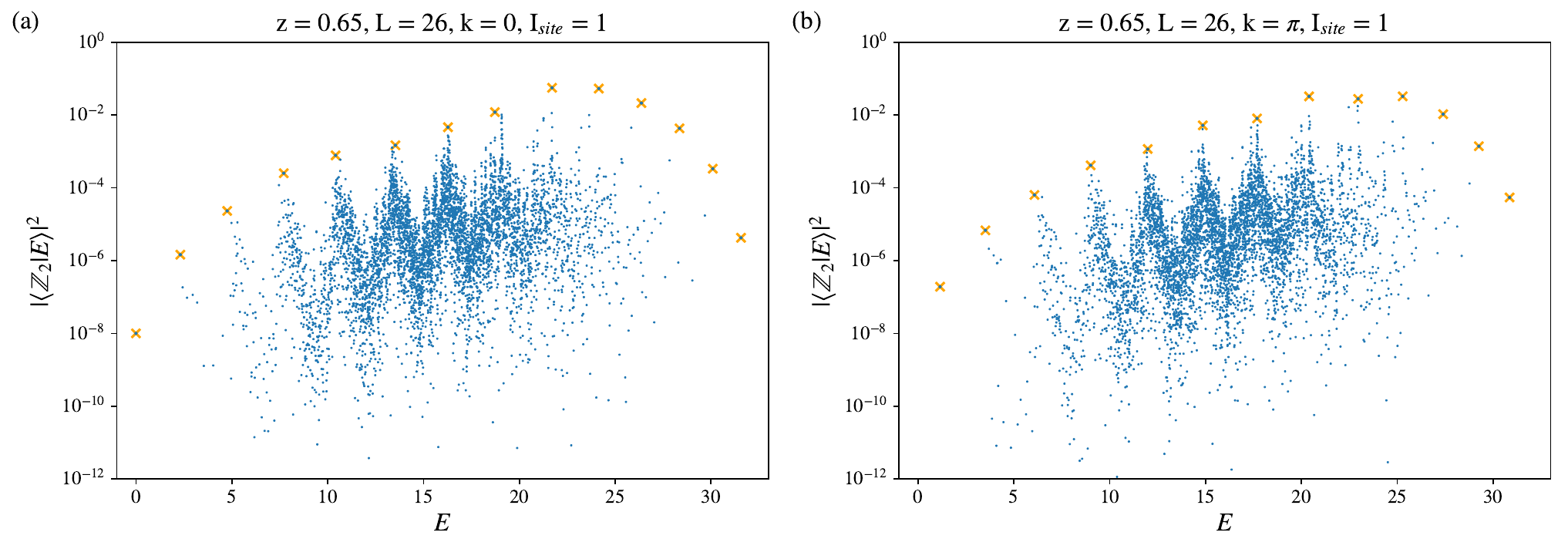}
\caption{Overlaps between the $\mathbb{Z}_2$ CDW and the eigenstates of the Lesanovsky model at $z=0.65$; the system size is $L=26$; the sector momentum quantum numbers are $k=0$ and $k=\pi$ in the left and right panels respectively, while the inversion quantum number is $I_\text{site} = 1$ in both panels. A band of states with high $\mathbb{Z}_2$ overlap is marked with crosses; along the energy axis, consecutive such scar states have momenta alternating between $k=0$ and $k=\pi$.
}
\label{fig:z 0.65 Z2 overlap}
\end{figure*}

\begin{figure*}[tb]
\centering
\includegraphics[width=\textwidth]{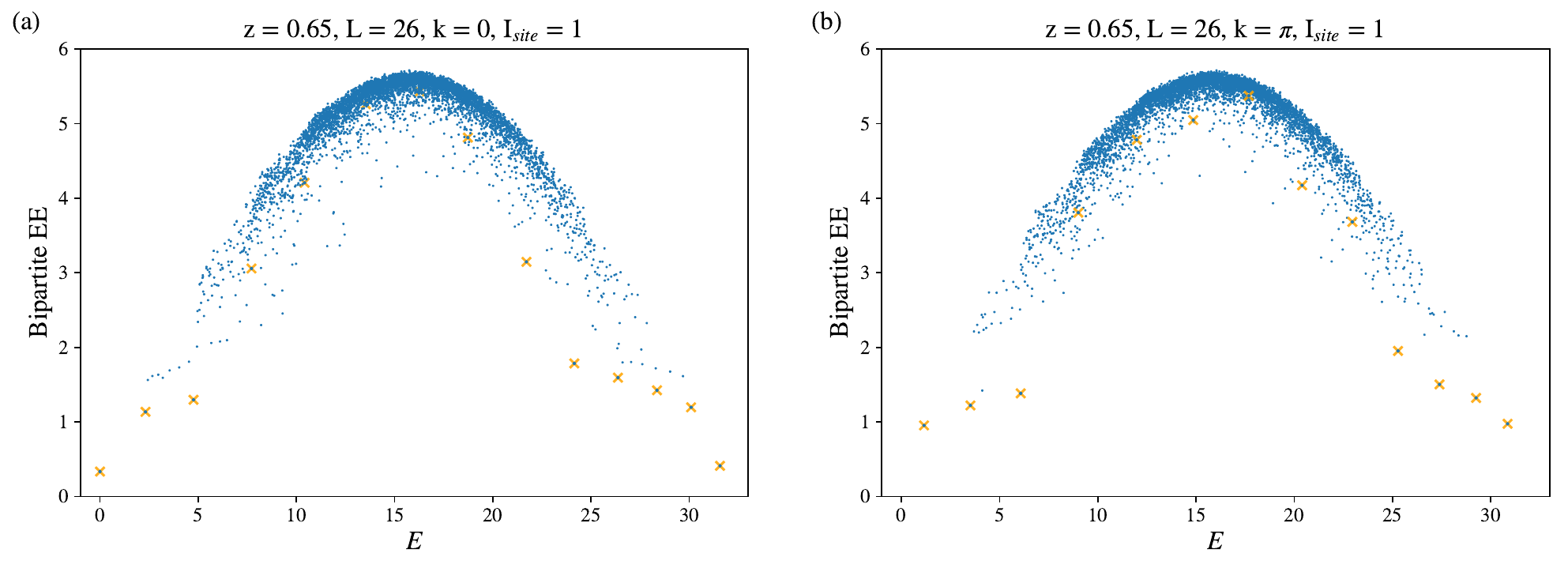}
\caption{Bipartite EE at $z=0.65$, $L=26$, for eigenstates in the symmetry sectors: (a) $k=0$ and (b) $k=\pi$; and 
$I_\text{site} = 1$ in both panels. States with high $\mathbb{Z}_2$ overlap identified in Fig.~\ref{fig:z 0.65 Z2 overlap} are marked with crosses.
}
\label{fig:z 0.65 EE}
\end{figure*}

\section{(Approximate) Connection to the PXP model at intermediate $z$} \label{section: PXP model}

\subsection{Ground state overlap}
\label{section:Ground state overlap}
The Lesanovsky model is particularly interesting in its possible connections to the PXP model. Numerically we observed that the Lesanovsky ground state $\ket{z}$ has the highest overlap squared with the PXP ground state at $z \approx 0.65$, with overlap squared of $\abs{\braket{gs_\text{PXP}}{z}}^2 = $ 0.977 at $L=26$.

This value of $z$ can be estimated analytically by evaluating the expectation value and variance of the operator $\sum_j P_{j-1} X_j P_{j+1}$ in the state $\ket{z}$. This is done by manipulating a classical dimer partition function~\cite{Lesanovsky2011, Lesanovsky2012}. We obtain:
\begin{equation} \label{eq:PXP expval}
\expval{\sum_j P_{j-1} X_j P_{j+1}}{z} = \frac{L}{z} \left(\frac{1}{\sqrt{1 + 4z^2}} - 1 \right) ~,
\end{equation}
and:
\begin{align} \label{eq:PXP variance}
&\expval{\big(\sum_j P_{j-1} X_j P_{j+1}\big)^2}{z} - \expval{\sum_j P_{j-1} X_j P_{j+1}}{z}^2 \\
&= \frac{L}{2} \left(1 + \frac{8}{(1 + 4z^2)^{3/2}} - \frac{17}{\sqrt{1 + 4z^2}} + 5\frac{1 + \sqrt{1 + 4z^2}}{z^2} \right) ~. \nonumber
\end{align}

The expectation value is minimized at $z = \frac{1}{2}\sqrt{\phi} \approx 0.6360$~\cite{Turner2018}, where $\phi$ is the golden ratio. The minimum value is $-0.6006 L$, close to the PXP ground state energy of $-0.6034 L$~\cite{Iadecola2019}. The variance attains a minimum of $0.0102 L$ at $z = 0.6205$. All these different ways to optimize proximity of the Lesanovsky ground state to the PXP ground state give values of $z$ that are close to each other.  In what follows, we will use $z = 0.65$ obtained by maximizing the ground state overlap.

\subsection{Approximate ``particle-hole symmetry'' of ground/ceiling state at $z=0.65$}
\label{section:Approximate PHS}
In addition to the high ground state overlap, we also noticed a high overlap between the ceiling states (the highest energy states) of the PXP model and the Lesanovsky model at $z=0.65$. We observed an overlap squared of 0.965 at $L=26$. We can understand this roughly as follows. The PXP ground state $\ket{\text{gs}_\text{PXP}}$ and the ceiling state $\ket{\text{cs}_\text{PXP}}$ are particle-hole partners: $\ket{\text{cs}_\text{PXP}} = \mathcal{C} \ket{\text{gs}_\text{PXP}}$, where $\mathcal{C} = \prod_{j=1}^L Z_j$.

For the Lesanovsky wavefunctions defined by Eq.~(\ref{eq:Ground state}), we have:
$\mathcal{C} \ket{z} = (-1)^L \ket{-z}$. Using

\begin{equation*}
\sum_j P_{j-1} X_j P_{j+1} \ket{z} = -\sum_j P_{j-1} (z^{-1} n_j + z P_j) P_{j+1} \ket{z} ~,    
\end{equation*}
we obtain:
\begin{equation}
H \ket{-z} = 2 \sum_j P_{j-1} X_j P_{j+1} \ket{-z} \approx 1.2012 L \ket{-z} ~.   
\end{equation}
Thus, the particle-hole partner $\ket{-z}$ is an approximate eigenstate of $H$ with trial energy $1.2012 L$, which compares well with the actual ceiling state energy $1.2140 L$ at $L=26$.

\subsection{Proximity to the PXP model}
\label{section:Proximity to PXP model}
We can formalize the proximity of the PXP model and the $z=0.65$ Lesanovsky model over the entire spectrum by arguing that the term $H_\text{classical} =\sum_j P_{j-1}  (z^{-1} n_j + z P_j) P_{j+1}$ constitutes a relatively small perturbation compared to the term $H_\text{PXP}$.  We can find the ground and ceiling states of $H_\text{classical}$ exactly.

For $z \leq 1/\sqrt{3}$, the classical ground state is $\ket{00...0}$, with energy $E_\text{classical} = z L$.
For $z > 1/\sqrt{3}$, the ground state is $\ket{\mathbb{Z}_3} = \ket{100100100...}$ with degeneracy 3 (or the closest strings to $\ket{\mathbb{Z}_3}$ if 3 does not divide $L$), with energy $(3z)^{-1} L$.

For $z \leq 1/\sqrt{2}$, the classical ceiling state is $\ket{\mathbb{Z}_2}$, with energy $(2z)^{-1} L$, and for $z > 1/\sqrt{2}$, the ceiling state is $\ket{00...0}$ with energy $z L$.

This gives the above operator $H_\text{classical}$ a spectral radius of $[(2z)^{-1} - (3z)^{-1}] L/2 = 0.1282 L$ at $z=0.65$. 

In comparison, the spectral radius of the PXP operator is $0.6034L$ \cite{Iadecola2019}. Given this, we can view the $z=0.65$ Lesanovsky model as a $\sim 20 \%$ deformation from the PXP model.

\subsection{Approximate scar states at $z \approx 0.65$}
\label{section:Scar states at z 0.65}
Given the proximity of the two models, we then studied the overlaps $\abs{\braket{\mathbb{Z}_2}{E}}^2$ between the $\mathbb{Z}_2$ CDW state and the $z=0.65$ eigenstates (Fig.~\ref{fig:z 0.65 Z2 overlap}) and the bipartite entanglement entropies (EE) of the eigenstates (Fig.~\ref{fig:z 0.65 EE}). The bipartite EE is a common measure of the violation of the strong ETH. In particular, while thermal eigenstates at fixed energy density are expected to scale with the volume of the system, in 1d, scar states in various systems are observed to have EEs $S\propto \log L$~\cite{turner_quantum_2018, moudgalya_entanglement_2018, schecter_weak_2019}, corresponding to sub-volume law entanglement. States with sub-volume law entanglement at finite energy densities do not obey the ETH, which
indicates a violation of the strong ETH. While the scarring in the Lesanovsky model is less prominent than in the PXP model, we observed a band of states with high $\mathbb{Z}_2$ overlap, see Fig.~\ref{fig:z 0.65 Z2 overlap}.  These states alternate between wavevectors $k=0$ and $k=\pi$, and have inversion number $I_\text{site} = 1$.

Marking out these states in the entanglement entropy plots (crosses in Fig.~\ref{fig:z 0.65 EE}) shows that near the edges of the spectrum, the states with high $\ket{\mathbb{Z}_2}$ overlap are also EE outliers. While numerical data with current system sizes is insufficient to conclusively show sub-volume law scaling, their unusual EE and $\ket{\mathbb{Z}_2}$ overlaps and their connection to integrable model states (Section~\ref{section:integrable ansatz}) suggests that they are ``approximate scar states." In contrast, the states in the middle of the spectrum have more typical EE values and their status as ``scar states" is more dubious.

\begin{figure*}
\centering
\includegraphics[width=\textwidth]{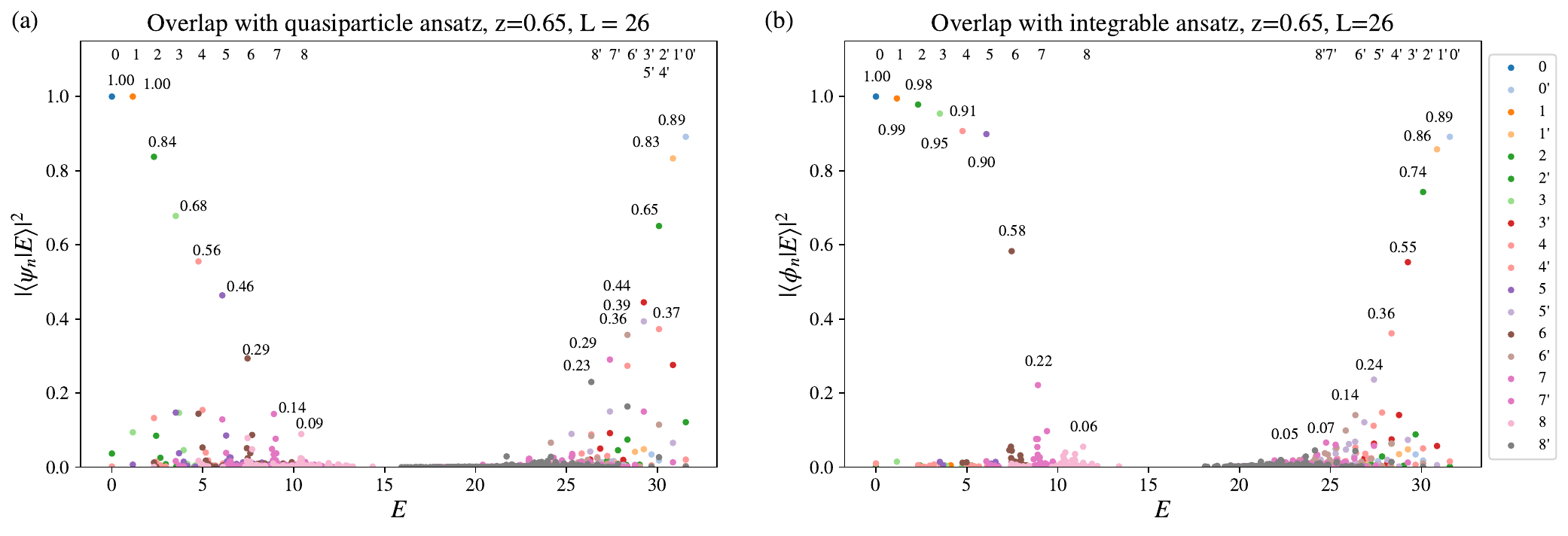}
\caption{Overlaps between the eigenstates of the Lesanovsky model at $z=0.65$ and (a) the quasiparticle $L_-^n$ ansatz $\ket{\psi_n}$ from Sec.~\ref{section:quasiparticle ansatz} and (b) the integrable model ansatz $\ket{\phi_n}$ from Sec.~\ref{section:integrable ansatz}; the system size is $L=26$. By construction the overlap is non-zero only for eigenstates in sectors with momentum $k=0$ (for even $n$) or $k=\pi$ (odd $n$) and inversion quantum number $I_\text{site} = 1$. We only displayed trial states $n \leq 8$, because further trial states have maximal overlaps $< 0.05$. For each $n$, the maximal overlap is indicated. The row of numbers along the top edge indicates the values $n$, their positions indicate the corresponding energy of the eigenstate with the highest overlap (the color of the corresponding highest overlap data point also identifies all data for given $n$). The numbers $n'$ indicate the charge conjugation of the $n^\text{th}$ trial state. Additionally, we verified that almost all the states with maximal overlap coincide with the $\mathbb{Z}_2$ outliers identified in Fig.~\ref{fig:z 0.65 Z2 overlap} (not shown).}
\label{fig:Quasiparticle ansatz}
\end{figure*}

\subsection{Approximation of scar states - quasiparticle ansatz}
\label{section:quasiparticle ansatz}
We define operators:
\begin{align}
    L_{\pm} &= \alpha_{\pm} \sum_{j=1}^L (-1)^j P_{j-1} \sigma^+_j P_{j+1} \nonumber \\
     & -z \beta \sum_{j=1}^L (-1)^j P_{j-2} \sigma^+_{j-1} P_j \sigma^+_{j+1} P_{j+2} ~,
\end{align}
where $\alpha_\pm$ and $\beta$ are defined in Eq.~(\ref{eq: Lesanovsky states equation}).
These are related to the exact Lesanovsky states by: $\ket{E_\pm} = L_\pm \ket{z}$.

Note that the writing here differs slightly from the original Lesanovsky expression in Eq.~(\ref{eq:Lesanovsky states}): we use $-z P\sigma^+_jP$ instead of $n_j$. The two operators coincide on $\ket{z}$.
However, the former is a more natural choice
for the multi-quasiparticle construction below in light of the rotated picture, cf. Appendix~\ref{section:Two quasiparticle}.

We expect to be able to superpose multiple Lesanovsky particles. Their mutual interactions are short-ranged and are associated with their contact.  We see this clearly and improve upon the two-quasiparticle ansatz in Appendix~\ref{section:Two quasiparticle}.  As $L \rightarrow \infty$, short-range interactions are rare, and we expect the quasiparticle ansatz to hold for finite number of particles.

We numerically test this and observe that while superposing multiple $L_-$ particles approximates ED eigenstates, superposing any number of $L_+$ particles produces states with moderate overlaps over multiple ED eigenstates.

We therefore focus on the following ansatz:
\begin{equation}
    \ket{\psi_n} = L_-^n \ket{z} ~, \quad n = 0, 1, \dots, L/2 ~.
    \label{eq:qpansatz}
\end{equation}
We repeat this construction until $\ket{\psi_{L/2}}$, which is the $k=0$ $\mathbb{Z}_2$ state if $L/2$ is even, or the $k=\pi$ $\mathbb{Z}_2$ state otherwise. For each $\ket{\psi_n}$, we evaluate the overlap squared $\abs{\braket{\psi_n}{E}}^2$ with every ED eigenstate $\ket{E}$. This is plotted in Fig.~\ref{fig:Quasiparticle ansatz}(a), where the maximum overlap for each $\ket{\psi_n}$ is also displayed, and eigenstates $\ket{E}$ with high $\mathbb{Z}_2$ overlap (Fig.~\ref{fig:z 0.65 Z2 overlap}) are marked with crosses. The trial states $\ket{\psi_n}$ are good descriptions of the identified scar states from $n=0$ to 5. They cease to be good descriptions for larger $n$. This accompanies the fact that the $\mathbb{Z}_2$ outliers in the middle of the spectrum cease to be entanglement entropy outliers (Fig.~\ref{fig:z 0.65 EE}). The quasiparticle picture evidently breaks down in this region, likely due to mixing with other states in the thermal bulk of the states.

Having previously identified approximate particle-hole symmetry of the ground and ceiling states, we then consider the charge conjugation of the above ansatz:
\begin{equation}
    \ket{\psi_{n'}} = \mathcal{C} L_-^n \ket{z} ~,~ n = 0, 1, \dots, L/2~.
\end{equation}
Again, the overlap squared with the ED eigenstates shown in Fig.~\ref{fig:Quasiparticle ansatz}(a) suggests that the $\ket{\psi_{n'}}$ are good scar state approximates for $n'=0'$ to $4'$, as seen by the fact that the ED eigenstates with high overlaps are predominantly marked with crosses.  Although the approximate particle-hole symmetry is not well understood, it appears to hold for the edges of the spectrum. For higher $n$, the eigenstates again lose their quasiparticle description. 

Notably, however, there is a set of EE outliers from $E \approx 22 - 26$ (at $L=26$) which are not captured by this ansatz. This is resolved by noting that the constructed trial states $\{\ket{\psi_n}\}, \{\ket{\psi_{n'}}\}$ are not orthogonal. We performed a Gram-Schmidt orthonormalization to produce an orthonormal basis $\{\ket{\Phi_i}\}_{i=0}^{L}$ and obtained
moderate overlaps $\sum_i \abs{\braket{s_m}{\Phi_i}}^2 \sim 0.5$ with the scar states $\ket{s_m}$ in this region (to keep the presentation simple, this analysis is not shown in the figure).

Following the measurement scheme of Ref.~\cite{Iadecola2019}, we also note that the subspace spanned by our ansatz states has, on average, overlaps squared of 0.47 with the identified scar state candidates. That is, $\sum_{m,i} \abs{\braket{s_m}{\Phi_i}}^2/(L+1) = 0.47$. This number is relatively low due to the poor overlaps in the middle of the spectrum. However, we note that this measure is not that far from the $\pi$-magnon tower construction in the PXP model in Ref.~\cite{Iadecola2019}, where such average overlap squared is 0.77 for $L=26$.

\subsection{Approximation of scar states - integrable ansatz}
\label{section:integrable ansatz}
We can refine our understanding of the scar states by approximating them with (properly rotated) eigenstates of the integrable hopping model $H_\text{eff.}^{(0)}$ in Eq.~(\ref{eq:H0eff}), which also happens to be an important part of the rotated Hamiltonian $H_R$, cf.\ the first line in Eq.~(\ref{eq: rotated Ham}).

Specifically, given a sector of fixed Rydberg excitation $n$, we find the integrable model eigenstate $\ket{\phi_n}_R$ of lowest energy in the symmetry sector $I_\text{site} = 1$ and $k=0$ or $k=\pi$ if $n$ is even or odd respectively. These states can be obtained from solving the relevant Bethe ansatz equations, see Ref.~\cite{Alcaraz1999}. We obtain the following ansatz for approximating the scar states of the Lesanovsky model:
\begin{align}
    \ket{\phi_n} &= S \ket{\phi_n}_R ~, \\
    \ket{\phi_n}_R &= \sum_{P \in S_n} \sum_{\{x_j\},\text{Ryd.}} A_P \exp(i \sum_{j=1}^n k_{P(j)} x_{j}) \ket{x_1 ... x_n}~,
\end{align}
where quasimomenta $k_j$ and amplitudes $A_P$ satisfy conditions discussed in Ref.~\cite{Alcaraz1999}. The sums are over all permutations $P$ of $n$ objects, and over all $n$-excitation states labelled by the positions of the excitations $1 \leq x_1 < ... < x_n \leq L$.

We choose the states that minimize the hopping energy $2 \sum_{j=1}^n \cos(k_j)$. Heuristically, these states will have quasimomenta $k_j$ as close as possible to $\pi$. These states are chosen because having quasimomenta $k_j \approx \pi$ minimizes connections to sectors with different excitation numbers, and we expect these states from the integrable model to persist also for the full Hamiltonian $H_R$ into regimes of moderate $z$. This is discussed in greater detail for the case of two quasiparticles in Appendix~\ref{section:Low energy integrable}. We refer to these states as the ``integrable ansatz."

For $n \leq 5$, the ansatz states $\ket{\phi_n}$ well approximate the ED scar states, with very high overlaps squared of over 0.9, see Fig.~\ref{fig:Quasiparticle ansatz}(b). The overlaps $\abs{\braket{\phi_n}{E}}^2$ sharply drop off for larger $n$. This is consistent with the sharp change in the entanglement entropy behavior of the `scar states' after $n=5$ in Fig.~\ref{fig:z 0.65 EE}. As with $\ket{\psi_n}$, we can charge conjugate the integrable ansatz to produce trial high energy states, denoted $\ket{\phi_{n'}} = \mathcal{C} \ket{\phi_{n}}$. These are slightly better approximates to the ED scar states than $\ket{\psi_{n'}}$ are, and their quality likewise sharply drops off at $n \approx 5$.

The marked improvement for $n \leq 5$ suggests that the physics of the quasiparticle interactions at low densities is well captured by the Bethe ansatz. The Bethe ansatz models the hard-core exclusion of the Rydberg excitations. Notably, it does not include the hybridization physics between `free' and `bound' magnons that the Lesanovsky quasiparticle ansatz $L_-^n$ includes. For example, $\ket{\phi_1} = \sum_{j=1}^L (-1)^j P_{j-1} \sigma^+_j P_{j+1} \ket{z}$, which does not contain the $\beta$ term present in the exact state $\ket{E_-}$ in Eq.~(\ref{eq:Lesanovsky states}).

These hybridization effects turn out to be small. $\ket{\phi_1}$ well approximates $\ket{E_-}$, with $\abs{\braket{\phi_1}{E_-}}^2 = 0.99$. This is because the ratio $\beta/\alpha_- = 0.20$ is relatively small, and the liquid state nature of $\ket{z}$ further reduces the effect of $\beta$, because the state from the $\beta$ term is largely parallel to that of the $\alpha_-$ term. A more refined trial state could include both repulsion interactions from the Bethe ansatz and hybridization effects from the Lesanovsky $\ket{E_-}$ state. 

\begin{figure}
\centering
\includegraphics[width=0.5\textwidth]{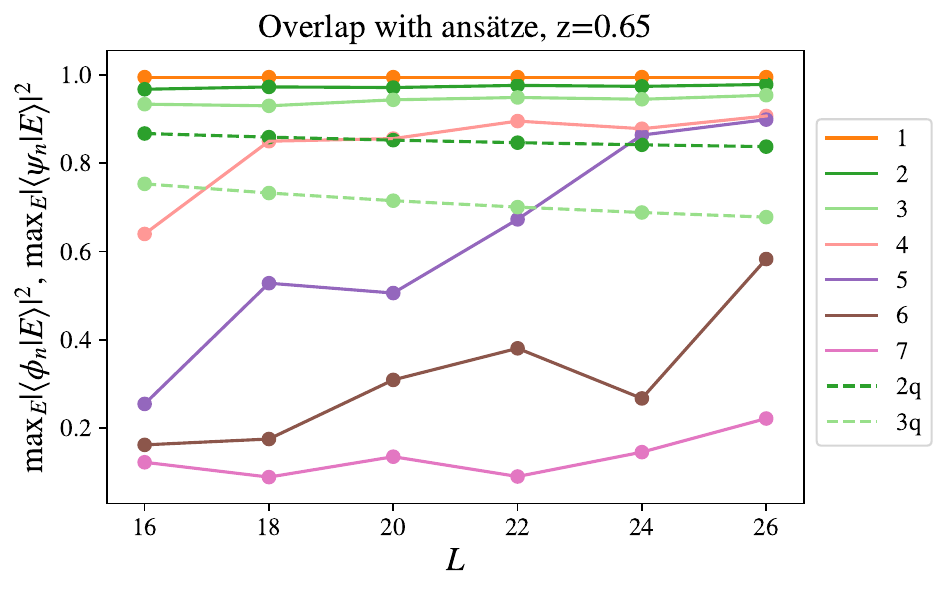}
\caption{Maximum overlap squared of each integrable ansatz $\ket{\phi_n}$ against system size $L$, for $n=1$ to $7$. The maximal overlap increases with $L$, suggesting that this ansatz remains good even in larger system sizes. We also plot with dashed lines the maximal overlaps for the quasiparticle ansatz $\ket{\psi_n}$ for $n=2,3$ ($n=1$ is exact since $\ket{\psi_1} = \ket{E_-}$). In this case the overlaps decrease with $L$.}
\label{fig:FiniteSize}
\end{figure}

To study the validity of our results for larger system sizes, we note that for fixed $n$, the overlaps between the ``integrable ansatz'' states and ED states increase with system size $L$, while the ED overlaps with the ``quasiparticle ansatz'' states decrease with $L$~(Fig.~\ref{fig:FiniteSize}), suggesting that the integrable ansatz remains a particularly good approximation even in larger system sizes. For fixed $n$, we expect the squared overlaps to saturate as $L$ increases, where for small $n$ the limiting values are apparently close to $1$.
The saturation is due to the fact that these are essentially finite-energy excitations constructed on top of the exactly known ground state (in particular, their energy density goes to zero as $L$ increases). We note, however, that the corresponding overlaps for the charge conjugated states $\ket{\phi_{n'}}$ decreases with $L$, indicating that the charge conjugation symmetry is only approximate.

\begin{figure}
\centering
\includegraphics[width=0.5\textwidth]{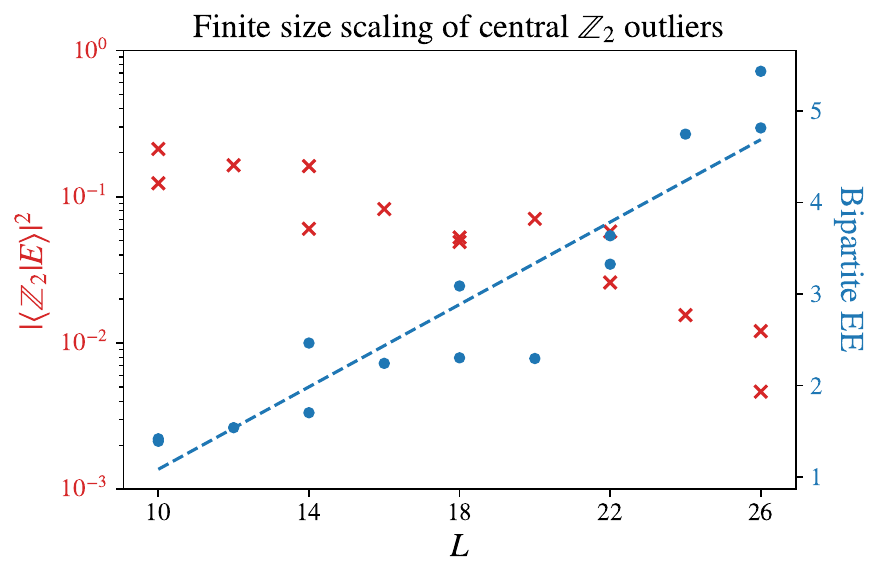}
\caption{Finite size scaling of central $\mathbb{Z}_2$ outlier states. At $k=0$, there are $L/2+1$ $\mathbb{Z}_2$ outlier states. If $L/2$ is even, we plot the $L/4$th $\mathbb{Z}_2$ outlier state, while if $L/2$ is odd, we plot the $(L-2)/4$th and $(L+2)/4$th states. These states have energy densities near $E/L = 0.68$. While the $\mathbb{Z}_2$ overlap (crosses) decreases roughly exponentially with $L$, the bipartite EE (circles) increases roughly linearly with $L$ (with linear fit plotted), suggesting volume-law EE scaling of $\mathbb{Z}_2$ outlier states at this energy density.}
\label{fig:FiniteSize2}
\end{figure}

In the middle of the spectrum, both approximations are not valid and we instead directly study the finite size scaling of the $\mathbb{Z}_2$ overlap and the bipartite entanglement entropy. We choose the $\mathbb{Z}_2$ outlier states in the middle of the spectrum as follows: At $k=0$, there are $L/2+1$ $\mathbb{Z}_2$ outlier states. If $L/2$ is even, we plot the $L/4$th $\mathbb{Z}_2$ outlier state, while if $L/2$ is odd, we plot the $(L-2)/4$th and $(L+2)/4$th states. These states consistently have energy densities near $E/L = 0.68$. In Fig.~\ref{fig:FiniteSize2} we plot the finite size scaling of these quantities. The overlap $\abs{\braket{\mathbb{Z}_2}{E}}^2$ decreases roughly exponentially with $L$, although the decay on these sizes is still relatively slow [roughly, $\sim \exp(-0.2 L)$ for these sizes vs. overlap-square decay of $\phi^{-L} = \exp(-0.48 L)$ expected for a thermal (random) state]. On the other hand, the bipartite EE increases roughly linearly, indicating that the mid-spectrum $\mathbb{Z}_2$ outlier states obey volume-law EE scaling and strongly mix with the thermal background.

\section{Conclusion}
\label{section:Conclusion}
We have discovered two new exact eigenstates of the Lesanovsky model. Additionally, we have discussed an exact non-Hermitian rotated Hamiltonian, in which the eigenstates are particularly simple, and a low $z$ effective model, both of which show proximity to an integrable model. Lastly, we have discussed connections between the PXP model and the Lesanovsky model, in particular at $z \approx 0.65$.
The Lesanovsky model at this $z$ also exhibits approximate scar states made prominent by their overlaps with the $\mathbb{Z}_2$ CDW state, although they are weaker than the scars in the PXP model, particularly in other ETH measures like the entanglement entropy.  We have constructed good approximations for the scars near the boundaries of the spectrum based on the quasiparticle ansatz with multiple Lesanovsky quasiparticles that is similar to the $\pi$-magnon ansatz introduced by Ref.~\cite{Iadecola2019} in the PXP model.
We have also shown that the so-called integrable ansatz performs even better, likely due to better treatment of interactions between the quasiparticles. Our work thus contributes to improved understanding of such Rydberg systems in general and their apparent scar states in particular.

It would be interesting to further study the relation between the Lesanovsky model $\mathbb{Z}_2$ outliers and the PXP scar states. In particular, our approximations in the Lesanovsky model resemble the `$\pi$-magnon' tower of Ref.~\cite{Iadecola2019}, in which the PXP scar states were approximated by a tower operator repeatedly acting on the PXP ground state. Given the similarity between the Lesanovsky and PXP models near the edges of the spectrum, it would be interesting to study how terms in the Lesanovsky model fully thermalize the middle of the spectrum.

\begin{acknowledgments}
We thank Alvaro Alhambra, Anushya Chandran, Manuel Endres, Timothy Hsieh, Rahul Nandkishore,
Zlatko Papi\'{c}, Tibor Rakovszky, Brenden Roberts, Maksym Serbyn, Brian Timar, Christopher Turner, and Christopher White for valuable discussions.
D.~M.\ acknowledges funding from the James C.~Whitney SURF Fellowship, Caltech Student-Faculty Programs. This work was also supported by National Science Foundation through Grant DMR-1619696.
C.-J.~L.\ acknowledges support from Perimeter Institute for Theoretical Physics.
Research at Perimeter Institute is supported in part by the Government of Canada through the Department of Innovation, Science and Economic Development Canada and by the Province of Ontario through the Ministry of Economic Development, Job Creation and Trade.

\end{acknowledgments}

\appendix

\section{Extensions of exact states near $k=\pi$}
\label{section: Momentum ansatz}
\begin{figure*}[htb!]
\centering
\includegraphics[width=\textwidth]{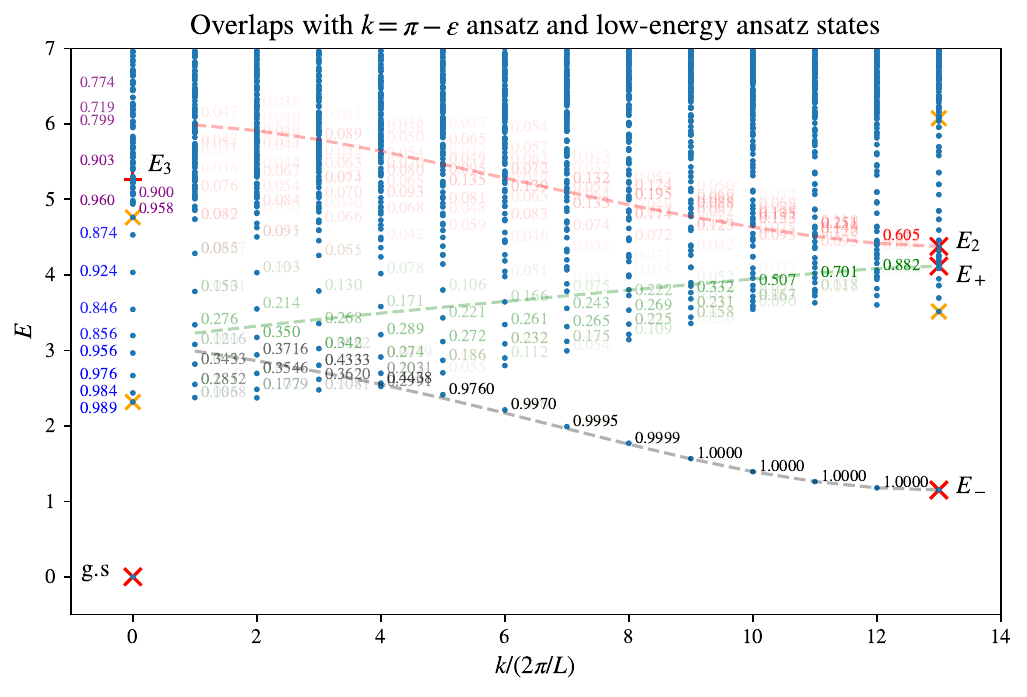}
\caption{Overlaps squared between the eigenstates of the Lesanovsky model at $z=0.65$ and the (a) $k=\pi-\epsilon$ ansatz states for $k = \pi/13$ to $k = 12\pi/13$ and (b) low-energy ansatz states; the system size is $L=26$. Overlaps of the $k=\pi-\epsilon$ ansatz (Appendix~\ref{section: Momentum ansatz}) corresponding to $\ket{E_-}, \ket{E_+}$ and $\ket{E_2}$ are marked in black, green and red respectively. Dashed lines are the eigenvalues of the $3\times3$ matrix formed by the connections between the three trial $k=\pi-\epsilon$ states discussed in Appendix~\ref{section: Momentum ansatz}.
Overlaps of the low energy ansatz (Appendix~\ref{section:Low energy integrable}) corresponding to the two-excitation ($k=0, I_\text{site} = 1$) and the three-excitation ($k=0, I_\text{site} = -1$) sectors are marked in blue and purple respectively. Each sector produces multiple trial states. For each trial state, the maximum overlap with an ED state is marked. For visibility, for states with $k=0$, only overlaps squared greater than 0.7 are shown.}
\label{fig:momentumansatz}
\end{figure*}

We use our understanding of the exact states $\ket{E_\pm}$ and $\ket{E_2}$ in the rotated frame $H_R$ of Eq.~(\ref{eq: rotated Ham}) to extend these states to wavevectors
\begin{equation}
k = \pi - \epsilon ~, \quad 0 < \epsilon \ll \pi ~.
\end{equation}
We study the $H_R$ matrix elements (`connections') between the following three states:

\scalebox{0.85}{
\begin{tikzpicture}
\node (010) [left] at (0,0) {$\sum_j e^{ikj}\ket{01_j0}$};
\node (101) [left] at (0,-3) {$\sum_je^{ikj}\ket{1_j01}$};
\node (1001) [left] at (5,-3) {$\sum_je^{ikj}\ket{1_j001}$};
\draw [->] (101.80) -- node[anchor=west] {$1+e^{i2\epsilon}$} (010.280);
\draw [->] (010.260) -- node[anchor=east] {$-z^2e^{-i\epsilon}$} (101.100) ;
\draw [->] (101.03) -- node[anchor=south] {$z(1-e^{-i\epsilon})$} (1001.177);
\draw [->] (1001.183) -- node[anchor=north] {$z(1-e^{i\epsilon})$} (101.357);
\draw [->] (1001.90) -- node[anchor=west] {$~1-e^{i3\epsilon}$} (010.0);
\node (dots1) [left] at (6.5,-3) {$...$};
\node (dots2) [left] at (6.5,-4) {$...$};
\draw [->] (1001.3) -- node[anchor=south]{$O(z \epsilon)$} (dots1.166);
\draw [->] (dots1.194) -- (1001.357);
\draw [->] (1001.345) -- node[anchor=west]{$~O(z^2 \epsilon)$} (dots2.150);

\path[<->] (010) edge [loop above] node {$z^{-1}+z(1-2\cos\epsilon)$} ();
\path[<->] (101) edge [loop below] node {$2(z^{-1}+z)$} ();
\path[<->] (1001) edge [loop below] node {$2(z^{-1}+z)$} ();
\end{tikzpicture}}

Here and in the later appendices, states are organized by excitation number; upward arrows are from the $\sigma^-$ term, downward arrows from the $P\sigma^+\sigma^-\sigma^+P$ term, horizontal arrows from the hopping term, and self-energies from both the hopping and chemical potential terms.

The above diagram gives a $3 \times 3$ matrix which can be diagonalized numerically. The three solutions correspond to trial states (in the rotated frame) for $k = \pi - \epsilon$ states evolving out of the $\ket{E_\pm}, \ket{E_2}$ states. We can then rotate back these states to obtain trial states in the original Lesanovsky frame.

The connections to additional states are only from the `1001' state and vanish linearly with $\epsilon$, and we expect good approximations of ED states near $\pi$ momentum. We indeed observe this in Fig.~\ref{fig:momentumansatz}, with very good overlaps for the $\ket{E_-}$ family of states, and good overlaps for the $\ket{E_+}$ and $\ket{E_2}$ trial states near $k=\pi$. The lowest energy trial ansatz in fact provides a very accurate description of the sharp quasiparticle branch outside of the two-particle continuum, which is also helped by the fact that the connections between the `010', `101' states and the `1001' state vanish linearly with $\epsilon$. The other two branches start inside the continuum and are sharp only for $k \to \pi$. The quick decline in the overlap for the $\ket{E_2}$ state is expected, because direct connections with states adjacent to the `1001' state (e.g., the `10001' state) were truncated.

Applying this approximation to systems with odd $L$, we obtain similar branches, and in particular obtain analogues of $\ket{E_\pm}$ and $\ket{E_2}$ in odd chains.

\section{Connection of low-energy states in the $k=0$ sector to integrable model states}
\label{section:Low energy integrable}
Here we present an approximation of low-lying states at $k=0$, including the two-quasiparticle state that belongs to the band of $\mathbb{Z}_2$ scars.

As in the previous section, we study the connections among a truncated set of states in the rotated frame $H_R$.  To model the two-quasiparticle state, we restrict our attention to states with excitation number zero, one, and two in the rotated frame.  We expect this to be a good approximation for the low energy states because of the high chemical potential energy $n (z^{-1} + z)$ of higher excitation numbers $n$.  The corresponding $k=0$ basis states and connections by the action of $H_R$ are displayed below:

\scalebox{0.7}{
\begin{tikzpicture}
\node (010) [left] at (0,0) {$\sum_j\ket{01_j0}$};
\node (101) [left] at (0,-3) {$\sum_j\ket{1_j01}$};
\node (1001) [left] at (3,-3) {$\sum_j\ket{1_j001}$};
\node (last) [left] at (10,-3) {$\sum_j\ket{1_j...1_{j+L/2}}$};
\node (dots) [left] at (5.5,-3) {$\cdots$};
\node(000) [left] at (3,2) {$\ket{00...0}$};
\draw [->] (101.80) -- node[anchor=west] {$2$} (010.280);
\draw [->] (010.260) -- node[anchor=east] {$z^2$} (101.100) ;
\draw [<->] (101.0) -- node[anchor=south] {$2z$} (1001.180);
\draw [->] (1001.90) -- node[anchor=south] {$2$} (010.310);
\draw [->] (last.182) -- node[anchor=north] {$4z$} (6,-3.05);
\draw [<-] (last.178) -- node[anchor=south] {$2z$} (6,-2.95);
\draw [->] (last.140) -- node[anchor=south] {$2$} (010.0);
\draw [<->] (1001.0) -- node[anchor=south] {$2z$} (4,-3);
\draw [->] (010.30) -- node[anchor=south] {$L$} (000.240);
\draw [->] (4.2,-2.7) -- node[anchor=south] {$2$} (010.330);
\draw [->] (5.8,-2.7) -- node[anchor=south] {$2$} (010.345);
\draw [gray,dashed,->] (1001.270) -- (1.9,-5.5);
\draw [gray,dashed,<-] (101.300) -- (1.7,-5.5);
\draw [gray,dashed,<-] (101.320) -- (3.8,-5.5);
\draw [gray,dashed,<-] (101.330) -- (5.8,-5.5);
\draw [gray,dashed,<-] (101.340) -- (8.3,-5.5);
\draw [gray,dashed,->] (2.1,-5.5) -- (4,-3.5);
\draw [gray,dashed,<->] (2.1,-5.6) -- (4,-5.6);
\draw [gray,dashed,->] (last.270) -- (8.5,-5.5);
\draw [gray,dashed,<-] (last.240) -- (6,-5.5);
\draw [gray,dashed,->] (4,-5.5) -- (1001.300);
\draw [gray,dashed,->] (4.1,-3.6) -- (4.1,-5.4);
\draw [gray,dashed,<->] (8.4,-5.6) -- (6.1,-5.6);
\draw [gray,dashed,->] (8.3,-5.4) -- (6.1,-3.5);
\draw [gray,dashed,<-] (6,-5.4) -- (6,-3.5);
\node(dots2) [left] at (5.5,-5.5) {$\cdots$};

\path[<->] (010) edge [loop above] node {$z^{-1}+3z$} ();
\path[<->] (101) edge [loop below] node {$2(z^{-1}+z)$} ();
\path[<->] (1001) edge [loop below] node {$~2(z^{-1}+z)$} ();
\path[<->] (last) edge [loop below] node {$2(z^{-1}+z)$} ();
\end{tikzpicture}}

Neglecting connections to higher excitation numbers yields a system of $L/2+1$ states. The connection with the state $\ket{00...0}$ is trivial because setting the coefficient $a_{00..0} = (L/\lambda) a_{010}$ satisfies the eigenvalue equation for any eigenvalue $\lambda$. We then obtain an $L/2 \times L/2$ matrix, which yields $L/2$ eigenvalues and eigenstate equations. The found eigenstates are easily rotated back into the Lesanovsky frame and compared to ED eigenstates. This is done in Fig.~\ref{fig:momentumansatz}, where only overlaps squared greater than 0.7 are marked in blue. The low-energy eigenstates of the $L/2 \times L/2$ matrix approximate well all states between the second to the third scar states (orange crosses) in the $k=0$ sector.

In particular, we notice that the matrix elements within the two-excitation sector are simply those of the integrable model in Ref.~\cite{Alcaraz1999}. We can view $L/2-1$ of the $L/2$ trial states as the integrable model states perturbed by the connections to the $\sum_j \ket{01_j0}$ state.

The lowest energy trial state, which approximates the two-quasiparticle state, experiences the smallest perturbation. This is because the particles have relative momentum close to $\pi$ (which can be viewed as individual particles carrying opposite quasimomenta close to $\pi$), and so this state has the weakest connection to the $\sum_j \ket{01_j0}$ state.  In fact, we find that both the trial state and the ED eigenstate in the rotated frame are well approximated by the integrable model solution:
\begin{align}
    \ket{\psi_\text{Int.}} &= \sum_{n = 2}^{L/2} c_n \sum_{j = 1}^L \ket{...01_j0...01_{j+n}0...} ~, \\
    c_n &= \cos\left[\pi \frac{L-3}{L-2} \left(n - \frac{L}{2} \right) \right] ~.
\end{align}
(The slight difference of the quasimomenta from $\pi$ in finite chains reflects the Bethe ansatz incorporation of the effects of interaction between the quasiparticles.)

We expect this property to hold with higher quasiparticle number: the quasimomenta near $\pi$ of the particles leads both to significant cancellations for connections to states with lower excitation number and to low energy under the hopping term, which minimizes mixing with states with higher excitation number.
While there will be some degree of mixing with these sectors, we expect them to mix in a similar way as in the Lesanovsky $\ket{E_-}$ excitation and only mix over a small part of the Hilbert space. These quasiparticle states then preserve more of their integrable model character, accounting for the observed low entanglement entropies in Fig.~\ref{fig:z 0.65 EE}. This inspires the `integrable ansatz' for approximating the scar states, which we study in Sec.~\ref{section:integrable ansatz}.

We also studied the three-excitation sector at $k=0$ and $I_\text{site} = -1$. Because there are no zero-, one-, and two-excitation states in this symmetry sector, we simply rotated the integrable model solutions in this sector back to the Lesanovsky frame and compared them to the ED eigenstates. The overlaps squared are marked in purple text in Fig.~\ref{fig:momentumansatz}.

Again, we find good overlaps for many states. However, states with energy near $E_3 = 3z^{-1} + z$ are not well approximated by the integrable model states, with only about $\sim 0.6$ overlap squared with the trial states. This is likely because the integrable model states are scattering states of three independent magnons, while the $\ket{E_3}$ state has two of the three magnons in a bound state. Nevertheless, we are able to well approximate other nearby states.

\section{Direct systematic improvement of the two-quasiparticle ansatz}
\label{section:Two quasiparticle}
Besides understanding the two-quasiparticle state through the integrable hopping model, we can directly improve on the two-quasiparticle ansatz $\ket{\psi_2} = L_-^2 \ket{z}$ through analysis in the rotated frame of Sec.~\ref{subsection: Proofs}. We define:
\begin{align}
    (P\sigma^+P)_\pi &\equiv \sum_{j=1}^L (-1)^j P_{j-1}\sigma^+_j P_{j+1} ~, \nonumber\\
    (P\sigma^+P\sigma^+P)_\pi &\equiv \sum_{j=1}^L (-1)^j P_{j-2} \sigma^+_{j-1} P_{j} \sigma^+_{j+1} P_{j+2} ~;
\end{align}
\begin{align}
    \ket{a} &\equiv (P\sigma^+P)^2_\pi \ket{0...0} ~, \nonumber\\
    \ket{b} &\equiv (P\sigma^+P)_\pi(P\sigma^+P\sigma^+P)_\pi \ket{0...0} ~, \\ 
    \ket{c} & \equiv (P\sigma^+P\sigma^+P)^2_\pi \ket{0...0} ~. \nonumber
\end{align}

We then study the action of $H_R$ in Eq.~(\ref{eq: rotated Ham}) on the states $\ket{a}$, $\ket{b}$, $\ket{c}$. These are graphically displayed below. 

The maps to additional states $\sum_j \ket{0_j10}$, $\sum_j \ket{1_j01}$, etc.,
are associated with ``contact interaction'' effects when two magnons (free or ``bound") touch. There are additional matrix elements between these states --- these connections, the states' self-energies, and subsequent connections
to further states are suppressed for readability.

\scalebox{0.85}{
\begin{tikzpicture}
\node (a) at (0,0) {$\ket{a}$};
\node (b) at (0,-2) {$\ket{b}$};
\node (c) at (0,-4) {$\ket{c}$};
\node (101) [left] at (3,0) {$\sum_j\ket{1_j01}$};
\node (1001) [left] at (5.2,0) {$\sum_j\ket{1_j001}$};
\node (10101) [left] at (3.05,-2) {$\sum_j\ket{1_j0101}$};
\node (100101) [align=left] at (4.45,-2){$\sum_j\ket{1_j00101}$\\ $+\ket{1_j01001}$};
\node (010) [left] at (3,1) {$\sum_j \ket{0_j10}$};
\draw [->] (a.280) -- node[anchor=west] {$2z^2$} (b.80);
\draw [->] (b.100) -- node[anchor=east] {$-2$} (a.260);
\draw [->] (b.280) -- node[anchor=west] {$z^2$} (c.80);
\draw [->] (c.100) -- node[anchor=east] {$-4$} (b.260);
\draw [->] (a.60) -- node[anchor=south] {$2$} (010.180);
\draw [->] (a.0) -- node[anchor=south] {$4z$} (101.180);
\draw [->] (b.70) -- node[anchor=south] {$1$} (101.240);
\draw [->] (b.60) -- node[anchor=south] {$-2$} (1001.270);
\draw [->] (b.0) -- node[anchor=south] {$-2z$} (10101.180);
\draw [->] (c.60) -- node[anchor=west] {$-4$} (10101.240);
\draw [->] (c.40) -- node[anchor=north] {$2$} (100101.270);
\draw [gray,dashed, ->] (010.20) -- (3.5,1.5);
\draw [gray, dotted, <->] (101.0) -- (1001.180);
\draw [gray, dashed, ->] (10101.04) -- (100101.176);
\draw [gray, dashed, ->] (100101.90) -- (101.310);
\draw [gray, dashed, <-] (10101.356) -- (100101.184);
\draw [gray, dashed, ->] (010.280) -- (101.80);
\draw [gray, dashed, ->] (101.100) -- (010.260);
\draw [gray, dashed, ->] (1001.90) -- (010.0);
\draw [gray, dashed, ->] (1001.300) -- (10101.30);
\draw [gray, dashed, ->] (10101.63) -- (101.270);
\draw [gray, dashed, ->] (100101.65) -- (1001.320);
\draw [gray, dashed, <->] (1001.0) -- (6.2,0);
\draw [gray, dashed, <->] (100101.0) -- (6.2,-2);
\draw [gray, dashed, ->] (100101.300) -- (6.2,-3.5);
\path[<->] (a) edge [loop left] node {$2(z^{-1}-z)$} ();
\path[<->] (b) edge [loop left] node {$3z^{-1}+z$} ();
\path[<->] (c) edge [loop left] node {$4(z^{-1}+z)$} ();
\end{tikzpicture}}

We can recover the ansatz $\ket{\psi_2} = L_-^2 \ket{z}$ by forming a $3 \times 3$ matrix of the connections between the states $\ket{a}$, $\ket{b}$, and $\ket{c}$, and ignoring connections with other states. Diagonalizing the resultant $3 \times 3$ matrix has 3 solutions, which are $L_-^2 \ket{0...0}$, $L_+ L_- \ket{0...0}$, and $L_+^2 \ket{0...0}$ with energies $2E_-$, $E_- + E_+$, and $2E_+$ respectively. In the unrotated frame, these states are simply $L_{\pm} L_{\pm'} \ket{z}$, because $\left[L_\pm, S \right] = 0$. Our numerical study finds that only the $2E_-$ solution provides a good approximation of an ED eigenstate, and only trial states obtained by repeated applications of $L_-$ are presented in the main text.

For two quasiparticles, we can approximately account for the contact interactions between the Lesanovsky particles, by treating the additional states including all connections among them shown in the diagram below. In other words, we treat this non-Hermitian diagonalization problem by writing a truncated Hamiltonian in the basis generated by the ``leakage" from the approximate states $\{\ket{a}, \ket{b}, \ket{c}\}$.
In addition, we include the connection to the top-most state $\ket{00...0}$ because it follows trivially from $\sum_j \ket{0_j10}$, and because after unrotation by $S$, it contributes significantly to every state in the product state basis. We numerically observe that its inclusion noticeably improves the approximation.

\scalebox{0.85}{
\begin{tikzpicture}
\node (010) at (0,2) {$\sum_j \ket{0_j10}$};
\node (101) at (0,0) {$\sum_j \ket{1_j01}$};
\node (1001) at (3.2,0) {$\sum_j \ket{1_j001}$};
\node (10101) [left] at (1.2,-2) {$\sum_j \ket{1_j0101}$};
\node (101001) [align=left] at (3.4,-2) {$\sum_j\ket{1_j00101}$\\ $+\ket{1_j01001}$};
\draw [->] (010.260) -- node[anchor=east] {$z^2$} (101.95);
\draw [->] (101.75) -- node[anchor=west] {$2$} (010.280);
\draw [->] (101.5) -- node[anchor=south] {$2z$} (1001.175);
\draw [->] (1001.185) -- node[anchor=north] {$2z$} (101.355);
\draw [->] (1001) -- node[anchor=south] {$2$} (010);
\draw [->] (1001) -- node[anchor=north] {$2z^2$} (10101);
\draw [<-] (101.270) -- node[anchor=east] {$2$} (10101.110);
\draw [<-] (101.300) -- node[anchor=south]{$2$} (101001.130);
\draw [<-] (1001.325) -- node[anchor=west] {$2$} (101001.62);
\draw [->] (10101.5) -- node[anchor=south] {$z$} (101001.175);
\draw [->] (101001.185) -- node[anchor=north] {$2z$} (10101.355);
\path[<->] (010) edge [loop above] node {$z^{-1}+3z$} ();
\path[<->] (101) edge [in = 158, out = 145,loop] node[anchor=south] {$2(z^{-1}+z)~~$} ();
\path[<->] (1001) edge [loop above] node {$2(z^{-1}+z)$} ();
\path[<->] (10101) edge [loop below] node {$3(z^{-1}+z)$} ();
\path[<->] (101001) edge [in = 260, out = 280,loop] node [anchor=north] {$3z^{-1}+4z$} ();
\draw [<->,dashed] (101001.0) -- (5.5,-2);
\draw [<->,dashed] (1001.0) -- (5.5,0);
\draw [->,dashed] (1001.350) -- (5.4,-1.8);
\draw [->,dashed] (101001.350) -- (5.4,-3);
\node (000) [right] at (2,3) {$\ket{00...0}$};
\draw [->] (010) -- node[anchor=south]{$L$} (000.200);
\draw [<-,red] (101.180) -- node[anchor=south] {$4zv_{a}$} (-2,0);
\draw [<-,red] (101001.200) -- node[anchor=north] {$~~2v_{c}$} (1.1,-3.3);
\draw [<-,red] (10101.180) -- node[anchor=south] {$-2zv_{b}$} (-2,-2);
\draw [<-,red] (10101.200) -- node[anchor=east] {$-4v_{c}~$} (-2,-3.3);
\draw [<-,red] (1001.260) -- node[anchor=west] {$-2v_{b}$} (2.1,-1);
\draw [<-,red] (101) -- node[anchor=north] {$v_{b}$} (-1.7,-1);
\draw [<-,red] (010) -- node[anchor=south] {$2v_{a}$} (-1.6,1.5);
\end{tikzpicture}}

Truncating the connections at those displayed in the figure, we notice that there is no feeding back into the $\{\ket{a}, \ket{b}, \ket{c}\}$ states, and hence the eigenvalue remains unchanged, as well as the amplitudes $v_a, v_b, v_c$. 
Accounting for these additional connections is then equivalent to solving the following linear equation in the $\{\ket{00...0}$, $\sum_j \ket{0_j10}$, $\sum_j \ket{1_j01}$, $\sum_j \ket{1_j001}$, $\sum_j \ket{1_j0101}$, $\sum_j (\ket{1_j01001} + \ket{1_j00101}) \}$ basis:
\begin{align}
 & 2E_- \ket{\psi'} =  \left(\begin{smallmatrix}
 0 \\
2v_{a}\\ 4zv_{a}+v_{b}\\  -2v_{b}\\ -2zv_{b}-4v_{c}\\ 2v_{c}\\  \end{smallmatrix} \right)\\ &+ \left(
\begin{smallmatrix}
0 & L & 0 & 0 & 0 & 0\\
0& z^{-1}+ 3 z & 2 & 2 & 0 & 0\\
0& z^2 & 2(z^{-1}+z) & 2z & 2 & 2\\
0& 0 & 2z& 2(z^{-1}+z) & 0 & 2\\
0& 0 & 0 & 2z^2 & 3(z^{-1}+z) & 2z\\
0& 0 & 0 & 0 & z& 3z^{-1}+4z \\
\end{smallmatrix} \right) \ket{\psi'}, \nonumber
\end{align}
where $v_i$ are the coefficients of $L^2_-\ket{0...0}$ in the $\{\ket{a}, \ket{b}, \ket{c}\}$ basis. This correction corresponds to a reduction in probability when the Lesanovsky particles are nearby and gives sizable improvement in the overlap squared with an exact eigenstate from 0.837 to 0.926 for $z=0.65$ at $L=26$.
We can systematically continue by including the states connected to those additional states considered in this section. This further increases the overlap squared to 0.960. 

%


\end{document}